\documentclass{article}
\usepackage[preprint]{spconf}

\toappear{ACCEPTED VERSION, 2023 IEEE International Conference on Acoustics, Speech and Signal Processing}

\usepackage{amsmath,graphicx}

\usepackage{amsfonts}
\usepackage{mathtools}

\usepackage{siunitx}
\usepackage{booktabs}
\usepackage{url}

\usepackage{bm}
\renewcommand{\vec}[1]{\bm{#1}}

\usepackage{amsthm}
\usepackage{thmtools}

\newcommand{\freal}{\mathbb{R}}

\newcommand{\fnat}{\mathbb{N}}
\newcommand{\fint}{\mathbb{Z}}

\usepackage{bbm}

\newcommand{\summary}[1]{\mathord{\sim}\{#1\}}

\newcommand{\lEsNO}{E_{\textnormal{s}} / N_0}

\newcommand{\argmaxM}[1]{\underset{#1}{\operatorname{arg\,max}}\;}

\DeclareMathOperator{\softmax}{softmax}

\newcommand{\idx}[1]{\textnormal{#1}}
\newcommand{\setX}{\mathcal{X}}

\newcommand{\options}{\mathcal{L}}

\title{Structural Optimization of Factor Graphs for Symbol Detection via Continuous Clustering and Machine Learning}
\name{Lukas Rapp, Luca Schmid, Andrej Rode, and Laurent Schmalen\thanks{This work has received funding in part from the European Research Council (ERC) under the European Union’s Horizon 2020 research and innovation programme (grant agreement No. 101001899) and in part from the German Federal Ministry of Education and Research (BMBF) within the project Open6GHub (grant agreement 16KISK010).}
\thanks{\copyright\ 2023 IEEE. Personal use of this material is permitted. Permission from IEEE must be obtained for all other uses, in any current or future media, including reprinting/republishing this material for advertising or promotional purposes, creating new collective works, for resale or redistribution to servers or lists, or reuse of any copyrighted component of this work in other works. Article DOI: 10.1109/ICASSP49357.2023.10096360}
}
\address{Communications Engineering Lab, Karlsruhe Institute of Technology (KIT), Karlsruhe, Germany}

\begin{document}
\ninept
\maketitle
\begin{abstract}\vspace*{-0.5ex}
We propose a novel method to optimize the structure of factor graphs for graph-based inference. As an example inference task, we consider symbol detection on linear inter-symbol interference channels.  The factor graph framework has the potential to yield low-complexity symbol detectors. However, the sum-product algorithm on cyclic factor graphs is suboptimal and its performance is highly sensitive to the underlying graph. Therefore, we optimize the structure of the underlying factor graphs in an end-to-end manner using machine learning. For that purpose, we transform the structural optimization into a clustering problem of low-degree factor nodes that incorporates the known channel model into the optimization. Furthermore, we study the combination of this approach with neural belief propagation, yielding near-maximum a posteriori symbol detection performance for specific channels.
\end{abstract}
\begin{keywords}
Factor Graphs, Machine Learning, Symbol Detection
\end{keywords}

\section{Introduction}\vspace*{-1ex}
Factor graphs are powerful frameworks to efficiently compute inference tasks. This makes them an important tool in communication engineering, where the essence of many tasks is based on statistical inference, e.g., decoding of channel codes or the Viterbi algorithm~\cite{loeligerIntroduction2004}.
A factor graph represents the factorization of a global function of multiple variables in a graphical way. This enables the calculation of the marginals of the function for inference by a message passing algorithm called sum-product algorithm (SPA) \cite{kschischangFactor2001}.

The performance of the inference over a factor graph depends highly on the underlying graph itself. Therefore, an interesting problem is the learning of factor graphs to improve the inference performance for a given task.
Shlezinger et al.~\cite{shlezingerLearned2022a} learn factor graph models to estimate distorted symbols by learning the factor nodes (FNs) with neural networks (NNs).
Another application where optimized factors graphs are useful is localization in robotics~\cite{salasSLAM2013}. 
In \cite{baikovitzGround2021}, the local functions of a factor graph are calculated by NNs using data of a ground penetrating radar to estimate the location of the radar.
Yi et al.~\cite{yiDifferentiable2021a} propose a general framework to optimize factor graphs for state estimation in the context of robotics. In contrast to the two previous works, the factor graph is optimized in an end-to-end fashion with respect to the inference performance by backpropagation~\cite[Sec.~6.5]{goodfellowDeepLearning2016}. 
All of these approaches only focus on the optimization of the FNs and use a fixed factor graph structure.

For factor graphs with cycles, however, the SPA performance heavily relies on the factor graph structure \cite{knollFixedPoints2018}. In many applications, factor graphs with cycles are inevitable to keep the inference complexity low. Since the factor graph structure of a given global function is not unique, it can be optimized to improve the SPA performance.
So far, to the best of our knowledge, research on learning the factor graph structure is rare. 
In \cite{abbeelLearning2006}, Abbeel et al. learn the structure and parameters of a factor graph simultaneously from observed data.
On the other hand, machine learning of other probabilistic graphical models \cite{kollerProbabilistic2009} like Bayesian networks and Markov random fields has been already intensively studied. An overview can be found in \cite[Sec.~3]{kollerProbabilistic2009} and \cite{vowelsYa2021}. 
Since Bayesian networks and Markov random fields can be transformed into factor graphs \cite{kschischangFactor2001}, factor graph learning can also be performed by learning one of these models followed by a transformation.
However, these approaches focus on the learning of an unknown probability distribution but do not perform an optimization with respect to the SPA performance.

In contrast, we focus on the learning of the factor graph structure with respect to the SPA performance in this paper. Therefore, we assume that the global function of the factor graph is already known and only its structure (i.e., the exact factorization) is subject to optimization. This assumption is often fulfilled in communications because good mathematical models exist \cite{heModelDriven2019}.
For optimization, we follow the approach of \cite{yiDifferentiable2021a} and learn the factor graph in an end-to-end manner.
Previous investigations have shown that there is a trade-off between the SPA performance and complexity \cite{schmidNeural2022}, therefore, we constrain the optimization to factor graphs with an upper bounded SPA complexity.
Our proposed approach is based on the clustering of FNs, a basic factor graph transformation to eliminate cycles in a factor graph \cite{kschischangFactor2001}.
To test our approach, we consider symbol detection on linear inter-symbol interference channels \cite{colavolpeApplication2005} using techniques of our previous work on this task \cite{schmidLowcomplexity2022}.

\section{Background}
\subsection{Factor Graphs and Marginalization}
Let \(f(\setX)\) be a global function that depends on the variables \(\setX \coloneqq \{x_1, \dots, x_n\}\), which can be factorized into
\begin{equation*}\label{eqn:factorization}
    f(\setX) = \prod_{i = 1}^K f_i(\setX_i),
\end{equation*}
where \(\setX_i \subset \setX\). A \emph{factor graph} is a bipartite graph that represents this factorization. It consists of variable nodes  (VNs) \(x_i\) representing the arguments \(x_i \in \setX\) of \(f\) and of FNs \(f_i\) representing the factors \(f_i(\setX_i)\) of the factorization. We will refer to these factors as \emph{local functions} of \(f\). An FN \(f_i\) is connected with a VN \(x_j\) by an undirected edge if and only if \(f_i(\setX_i)\) depends on \(x_j\): \({x_j \in \setX_i}\)~\cite{kschischangFactor2001}.

Factor graphs enable an efficient approximation of the marginals \(f(x_i) = \sum_{\summary{x_i}} f(\setX)\) of a function \(f\) via the SPA \cite{kschischangFactor2001}.
The notation ``\(\sum_{\summary{x_i}}\)'' denotes the combined summations \(\sum_{x_j \in \mathcal{D}_j}\) over all variables except \(x_i\), where \(\mathcal{D}_j\) is the finite set of values that \(x_j\) can take on.
This work uses the SPA implementation of \cite{schmidLowcomplexity2022}, in which the specific update equations can be found:
First, the outgoing messages of each VN \(x_i\) are initialized by vectors with the entries \(1/|\mathcal{D}_i|\).
Then a flooding update schedule is applied, i.e., one iteration consists of a simultaneous update of all FNs followed by a simultaneous update of all VNs.
After \(N\) update iterations, the estimated marginals \(\hat{f}(x_i)\) are calculated with the final messages. 

For a factor graph with cycles, the estimated marginals only approximate the real marginals.
Kschischang et al.~\cite{kschischangFactor2001} present several transformations of factor graphs to modify its structure into a more convenient one. We will focus on \emph{clustering} of FNs in this paper.
To cluster two FNs \(f_1(\setX_1)\) and \(f_2(\setX_2)\), both FNs are replaced by a new FN \(f_{\idx{n}}(\setX_{\idx{n}})\) that is connected with all variables \(\setX_{\idx{n}} \coloneqq \setX_1 \cup \setX_2\) with which FN \(f_1\) or \(f_2\) have been connected. The local function of \(f_{\idx{n}}\) is the product of both local functions:
\(
    f_\idx{n}(\setX_{\idx{n}}) = 
    f_1(\setX_1)
    f_2(\setX_2)
\).

\subsection{Symbol Detection}
As an example inference task, we consider symbol detection~\cite{colavolpeApplication2005}, i.e., the estimation of symbols \(x_i\) from an observed noisy sequence \(\vec{y}\).
We assume that \(K\) independent and uniformly distributed binary phase-shift keying (BPSK) symbols \({\vec{x} \in \{\pm 1\}^K}\) are transmitted over a baseband channel. The channel distorts the sequence by linear inter-symbol interference described by the finite channel impulse response \(\vec{h} \in \freal^{L+1}\) and additive white Gaussian noise (AWGN). Instead of a standard convolution for the interference, we use a cyclic convolution, resulting in the following channel model~\cite{colavolpeApplication2005}:%
\footnote{%
We use the cyclic instead of the standard non-cyclic channel model to avoid boundary effects, 
which falsify the symbol detector performance and make a comparison between different models complicated.
In practice, however, the same methods can be applied on a non-cyclic channel.
}
\begin{equation}\label{eqn:channel-model}
    y_k = \sum_{\ell = 0}^{L} h_\ell x_{[k - \ell]_K} + w_k, \qquad k = 1, \dots, K.
\end{equation}
Here, \(w_k \sim \mathcal{N}(0, \sigma^2)\) is an AWGN sample with noise variance \(\sigma^2 = (2 \lEsNO)^{-1}\) and
\([a]_K \coloneqq a \mathbin{\textnormal{mod}} K\) denotes the modulo operation, which calculates the remainder of the division of \(a \in \fint\) by \(K \in \fnat\).
For performance evaluation, we consider the channel
\(
    \vec{h} = [\num{0.407}, \num{0.100}, \num{0.815}, \num{0.100}, \num{0.407}]
\) as an example.

In the following, we briefly review symbol detection based on the factor graph framework. For a more detailed introduction, we refer the reader to~\cite{schmidLowcomplexity2022, colavolpeSISO2011a}.
Symbol-wise maximum a posteriori (MAP) detection is given by
\begin{equation}\label{eqn:symbol-wise-map-rule}
    \hat{x}_k =
    \argmaxM{x_k \in \{\pm 1\}}
    P(x_k \mid \vec{y}),
    \quad
    P(x_k \mid \vec{y}) 
    = \sum_{\summary{x_k}} 
    P(\vec{x} \mid \vec{y}),
\end{equation}
for \(k = 1, \dots, K\). Using Bayes' theorem and \eqref{eqn:channel-model}, we obtain~\cite{colavolpeApplication2005}
\begin{align}
    P(\vec{x} \mid \vec{y}) 
    \propto \prod_{k=1}^{K}
    \underbrace{
        \exp 
        \Bigg(-\frac{1}{2 \sigma^{2}}
            \Big|
                y_{k}-\sum_{\ell=0}^{L} h_{\ell} 
                x_{[k - \ell]_K}
            \Big|^{2}
        \Bigg)
    }_{\coloneqq g(x_{[k - L]_K}, \dots, x_k)}.
    \label{eqn:factorization-channel-model}
\end{align}

The marginalization in \eqref{eqn:symbol-wise-map-rule} can be efficiently calculated with the factor graph framework resulting in the estimated marginals \({\hat{P}(x_k \mid \vec{y})}\) and the MAP estimates \(\hat{x}_k\).
The factor graphs of \eqref{eqn:symbol-wise-map-rule} are in general not cycle-free, causing the marginals to not be exact. 
Previously, we analyzed two factor graph models~\cite{schmidNeural2022}, which were first introduced in~\cite{colavolpeApplication2005, colavolpeSISO2011a}:
In the first model, each factor \(g(\cdot)\) in \eqref{eqn:factorization-channel-model} is associated with one FN~\cite{colavolpeApplication2005}. We refer to this graph as Forney-based factor graph (FFG). The second model~\cite{colavolpeSISO2011a} is given by
\begin{align}
    P(\vec{x} \mid \vec{y})
    &\propto 
    \prod_{k=1}^{K}\Bigg(
        F_k(x_k)
        \prod_{\ell = 1}^{L} 
        I_\ell(x_k, x_{[k + \ell]_K})
    \Bigg). \label{eqn:UFG}\\
    \intertext{This graph consists of the FNs \(F_k\) and \(I_\ell\)
    , which are given by}
    F_k(x_k) 
    &= \exp\left[
        \frac{1}{\sigma^2}
        \left(
            \left(\sum_{\ell=0}^L 
                h_\ell y_{[k + \ell]_K}
            \right) 
                x_k
            - \frac{q_0}{2} x_k^2
        \right)
    \right], \nonumber\\
    I_\ell(x_k, x_{[k + \ell]_K}) 
    &= 
    \exp\left(
        -\frac{1}{\sigma^2} q_\ell x_k x_{[k + \ell]_K}
    \right)\!, q_\ell
    = \sum_{i = 0}^{L}
    h_{i}
    h_{[i + \ell]_K},
    \nonumber
\end{align}
for \(k = 1, \dots, K\). For \(I_{\ell}\), we have \(\ell = 0, \dots, L\) and for \(q_{\ell}\), \(\ell = 1, \dots, L\).
We will refer to this graph as Ungerboeck-based factor graph (UFG).

The complexity of the SPA on a factor graph can be estimated by the number of message updates at the FNs \cite{colavolpeApplication2005}. An FN of degree \(d\) contributes a term proportional to \(2^d N\) to the overall complexity.
For low-complexity applications, factor graphs with a low maximum FN degree are thus required.
In terms of complexity, the UFG is preferred over the FFG because its maximum FN degree is \(2\) instead of \(L + 1\) for the FFG~\cite{schmidNeural2022}.
However, the UFG performs very poorly on the reference channel (see Fig.~\ref{plot:results} below) and many other channels whereas the FFG achieves near-MAP performance.
In the following, we propose methods to find factor graphs with a good performance while keeping the complexity low. For this, we limit the maximum FN degree \(d_{\idx{max}}\) of the learned factor graph to \(3\) or \(4\) in contrast to \(5\) for the FFG.

\section{Factor Graph Optimization}
Learning factor graphs from scratch involves different sub tasks that have to be jointly solved. This involves learning the VNs, FNs and edges connecting them.
In this work, we do not focus on learning of the VNs and associate each symbol \(x_k\) with one VN.
Instead, we focus on the graph structure and FNs because the complexity of the SPA primarily depends on them.
However, more complex VNs are possible, which, for instance, combine multiple symbols \cite[Sec. VI]{kschischangFactor2001}. Including the VNs in the optimization is subject of future work.

For the FNs \(f_i\), we must learn their local functions \(f_i(\setX_i, \vec{y})\), which also depend on the received symbols \(\vec{y}\) in addition to \(\setX_i\). 
One approach is learning these functions by NNs.
Let \(N_{\idx{FN}}\) be the number of FNs in the graph.
Assuming the SPA is accurate, the local functions must fulfill
\begin{equation}\label{eqn:constraint-fns}
    \prod_{i = 1}^{N_{\idx{FN}}}
    f_i(\setX_i, \vec{y})
    \stackrel{!}{\propto} P(\vec{x} \mid \vec{y}),
\end{equation}
so that the estimated marginals \({\hat{P}(x_i \mid \vec{y} )}\) match the correct mar\-gin\-als \({P(x_i \mid \vec{y})}\). Hence, the NNs indirectly need to relearn the known channel model \eqref{eqn:factorization-channel-model} every time the factor graph structure changes during the optimization.
This results in long training times and the need for large datasets.
We avoid this issue by limiting the search space to factor graphs that already fulfill \eqref{eqn:constraint-fns}.

We select a factor graph of the channel model \eqref{eqn:factorization-channel-model} called \emph{basis factor graph} (BFG) as the starting point of the optimization from which new graphs are created by clustering FNs.
These factor clustered graphs fulfill \eqref{eqn:constraint-fns} by construction.
The BFG should consists of many FNs of small degree resulting in a large search space of clustered factor graphs with small maximum FN degree.
In the following, we choose the UFG \eqref{eqn:UFG} with maximum FN degree \(2\) as BFG.
The motivation for this choice is that the FFG \eqref{eqn:factorization-channel-model} is a clustered version of the UFG.
Hence, we expect that a clustering of the UFG interpolates the performance and the complexity between the UFG and the FFG.

\subsection{Factor Node Containers}
\begin{figure}
    \centering
    \includegraphics{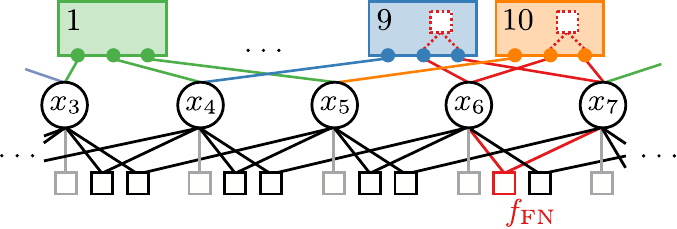}
    \caption{Concept of FN containers: The FNs on the bottom represent the BFG (here: UFG). The colored FNs above are the FN containers. 
    The red FN \(f_{\idx{FN}}\) can be clustered into container \(9\) and \(10\) (dashed) because \(f_{\idx{FN}}\) and both containers are connected with VN \(x_6\) and \(x_7\). Hence, its list of clustering options is: \(\options(f_\idx{FN}) = [9, 10]\).}
    \label{fig:fn-container}
    \vspace{-0.4cm}
\end{figure}

We introduce the concept of \emph{FN containers} to bring the clustering transformation into a systematic form that can be easily optimized. This concept is visualized in Fig.~\ref{fig:fn-container} by an example.
An \emph{FN container} \(f_{\text{C}}(\setX_{\text{C}})\) is a special FN that is connected to the VNs \(x_i \in \setX_{\text{C}}\) of a BFG and whose local function is  \(f_{\text{C}}(\setX_{\text{C}})=1\) before clustering.
Thereby, the FN container does not influence the SPA and can be considered as \emph{empty}.
We distinguish normal FNs and containers in the following by \(f_{\idx{FN}}\) and \(f_{\idx{C}}\).

These containers represent options into which the FNs of the BFG can be clustered: An FN \(f_{\idx{FN}}(\setX_{\idx{FN}})\) can be clustered into a FN container \(f_{\idx{C}}(\setX_{\idx{C}})\) if \(f_{\idx{C}}\) is connected to the same VNs as \(f_{\idx{FN}}\), i.e., \(\setX_{\idx{FN}} \subset \setX_{\idx{C}}\). When an FN \(f_{\idx{FN}}\) is clustered into a container \(f_{\idx{C}}\), \(f_{\idx{FN}}\) is removed from the factor graph and its local function is appended to the local function of the container by multiplication:
\(
    {f_{\idx{C}}(\setX_{\idx{C}}) \gets 
    f_{\idx{C}}(\setX_{\idx{C}})
    f_{\idx{FN}}(\setX_{\idx{FN}})}
\).

For a given BFG and a set of FN containers, we assign an ordered list \(\options(f_{\idx{FN}, j})\) of \emph{clustering options} to each FN \(f_{\idx{FN}, j}\). This list contains the indices of each factor node container in which \(f_{\idx{FN}, j}\) can be clustered:
\[
    \options(f_{\idx{FN}, j}) \coloneqq (1 \leq i \leq N_{\idx{C}} : \setX_{\idx{FN}, j} \subset \setX_{\idx{C}, i}), \quad \text{for \(1 \leq j \leq N_{\idx{FN}}\)},
\]
where \(N_{\idx{C}}\) is the number of FN containers. Note that we denote with round brackets an ordered list whose entries are arranged in ascending order.

\subsection{Factor Node Clustering}\label{sec:DC}
In this section, we motivate the basic concept of the clustering approach which is generalized in the next section.
We construct new factor graphs from the BFG by clustering each FN \(f_{\idx{FN}, i}\) into one of its possible containers \(f_{\idx{C}, j} \in \options(f_{\idx{FN}, i})\).
The filled FN containers together with the VNs form a new clustered factor graph. 
To find good factor graphs, the choices of the clustering options can be optimized with respect to the BER performance. 

After the optimization, each FN container \(f_{\idx{C}, i}\) can be simplified by checking on which variables its local functions \(f_{\idx{C}, i}(\setX_{\idx{C}, i})\) actually depend. The edges incident to variables on which the container does not depend can be removed because they have no impact on the result of the SPA.
This reduces the complexity of the SPA.
Empty FN containers, in which no FNs were clustered, end up with no connections to VNs and can be completely removed.

The performance of the optimized factor graph relies on the choice of FN containers. We limit the maximum degree \(d_{\idx{max}}\) of the learned factor graph by only using containers of degree \(d_{\idx{max}}\). Because of the edge removal after the optimization, FNs of smaller degree can still be learned.
Using all possible containers of degree \(d_{\idx{max}}\) is infeasible because of the large number of containers.
However, in our example, the dependency of VNs  \(x_i\) and \(x_j\) that are far apart, i.e., where \([i - j]_K\) is large, is negligible due to the finite impulse response \(\vec{h}\). Hence, containers connecting VNs that are far apart do not seem useful.
Motivated by this fact, we define the \emph{span} of an FN container \(f_{\idx{C}}(\setX_{\idx{C}})\) as the distance between its first and last VN: \(\max_{x_i, x_j \in \setX_{\idx{C}}} [i - j]_K\). Simulations have shown that a limitation to a span of \(L+1\) leads to good results and larger spans do not significantly improve the performance. Hence, in the following, containers are of degree \(d_{\idx{max}} = 3\) or \(4\) with span smaller or equal to \(L+1\).

\subsection{Continuous Clustering (CC)}
The clustering approach has the disadvantage that each FN is assigned to only one container, which limits the search space.
We can increase the search space by using every FN \(f_{\idx{FN}, i}\) simultaneously in all of its clustering options \(f_{\idx{C}, k} \in \options(f_{\idx{FN}, i})\). To achieve this, its local function \(f_{\idx{FN}, i}(\setX_{\idx{FN}, i})\) is factorized using the weights \(\alpha_{ij} \in \freal\):
\begin{equation}\label{eqn:cc-factorization}
    f_{\idx{FN}, i}(\setX_{\idx{FN}, i}) = 
    \prod_{j = 1}^{\mathclap{|\options(f_{\idx{FN}, i})|}}
    f_{\idx{FN}, i}^{\alpha_{ij}}(\setX_{\idx{FN}, i})
    \qquad
    \text{with}
    \quad
    \sum_{j = 1}^{\mathclap{|\options(f_{\idx{FN}, i})|}} 
    \alpha_{ij} 
    = 1,
\end{equation}
for \(i = 1, \dots, N_{\text{FNs}}\).
Thereby, each FN \(f_{\idx{FN}, i}\) can be clustered into each container \(f_{\idx{C}, k} \in \options(f_{\idx{FN}, i})\) using the factor \(f_{\idx{FN}, i}^{\alpha_{ik}}(\setX_{\idx{FN}, i})\). The local function of the \(m\)-th container is given by
\begin{equation}
    f_{\idx{C}, m}(\setX_{\idx{C}, m})
    = \prod_{(i, j) \in \mathcal{M}_m} 
    f_{\idx{FN}, i}^{\alpha_{ij}}(\setX_{\idx{FN}, i}),
\end{equation}
with
\(
    \mathcal{M}_m \coloneqq \{
        (i, j) : 
        1 \leq i \leq N_{\idx{FN}}, 
        1 \leq j \leq |\options(f_{\idx{FN}, i})|, 
        m = \options(f_{\idx{FN}, i})_j
    \}
\), where \(\options(f_{\idx{FN}, i})_j\) is the \(j\)-th entry of the list.

A valid factorization fulfilling the constraints 
\(
    \sum_{j = 1}^{|\options(f_{\idx{FN}, i})|} \alpha_{ij} = 1
\) in \eqref{eqn:cc-factorization} can theoretically be achieved with a combination of positive and negative exponents. However, our investigations have shown that negative exponents lead to poor results. Hence, we limit the weights to values between \(0\) and \(1\), which are given by \cite[Sec.~6.2.2.3]{goodfellowDeepLearning2016}:
\begin{equation}\label{eqn:softmax}
    \alpha_{ij} \coloneqq \softmax(\vec{\beta}_{i})_j
    \coloneqq \exp(\beta_{ij}) / \sum_{k = 1}^{\mathclap{|\options(f_{\idx{FN}, i})|}} \exp(\beta_{ik}),
\end{equation}
for \(i = 1, \dots, N_{\idx{FN}}\) and \(j = 1, \dots, |\options(f_{\idx{FN}, i})|\). 
The factor graph is parametrized by
\(
    {\vec{\beta}_i \in \freal^{|\options(f_{\idx{FN}, i})|}}
\) with
\(i \in \{1, \dots, N_{\idx{FN}}\}\), which need to be optimized.
Because of the softmax-function, the constraint in \eqref{eqn:cc-factorization} is automatically fulfilled.
The exponent \(\alpha_{ij} \in [0, 1]\) can be interpreted as the fraction with which the FN \(f_{\idx{FN}, i}\) is clustered into container \(f_{\idx{C}, j}\).
Note that the special case \(\alpha_{ij} \in \{0, 1\}\) for all exponents corresponds to the discrete clustering in Sec.~\ref{sec:DC}.

The weights \(\vec{\beta}_i\) are jointly optimized towards an objective function in an end-to-end manner using the Adam algorithm \cite{kingmaAdam2017} and backpropagation \cite[Sec.~6.5]{goodfellowDeepLearning2016} with learning rate \num{e-4}.
The source code for the factor graph optimization is available online \cite{sourceCode}.
The weights \(\vec{\beta}_i \) are initialized with random i.i.d. samples from a normal distribution \(\mathcal{N}(0, 1)\).
In each training step, a minibatch of \(D\) sequences  
\(
    \vec{x}^{(i)} \in \{\pm 1\}^K
\) 
with \(i = 1, \dots, D\) is generated, where each symbol is sampled randomly and independently from a uniform distribution.
Then, the channel model \eqref{eqn:channel-model} with \(\lEsNO = \SI{10}{\dB}\) is applied to all sequences resulting in \(\vec{y}^{(i)}\) for \(i = 1, \dots, D\)~\cite{schmidLowcomplexity2022}.
The estimated marginals 
\(m_{i, k} \coloneqq \hat{P}\big(x_k^{(i)} = +1 \big| \vec{y}^{(i)}\big)\) 
from the SPA for \(i = 1, \dots, D\) and \(k =1, \dots, K\) are evaluated with the soft BER \cite{lianLearned2019}
\[
    F(\vec{\beta}_1, \dots, \vec{\beta}_{N_{\idx{FN}}}) =
    \sum_{i = 1}^D \sum_{k = 1}^K
    m_{i, k}
    ^{\frac{1}{2}\left(1 - x_k^{(i)}\right)}
    (1 - m_{i, k})
    ^{\frac{1}{2}\left(1 + x_k^{(i)}\right)},
\]
which is used as objective function. Our simulation results have shown that the BER performance of the symbol-detection is significantly better if the underlying factor-graphs have been optimized with the soft BER instead of the cross-entropy.

After the optimization, the simplification rules for removing edges and FNs described in Sec.~\ref{sec:DC} cannot directly be applied because the exponents \(\alpha_{ij}\) are usually not exactly zero and hence, each FN container depends on all its variables. However, since FN components with \(\alpha_{ij} = 0\) have no impact on the SPA, we can assume that the impact of components with sufficiently small \(\alpha_{ij}\) on the SPA is negligible.
Because of that, we prune all FN components whose exponent \(\alpha_{ij}\) is below a certain threshold \(\alpha_{\idx{thr}}\) by setting 
\(\beta_{ij} = -\infty\) resulting in \(\alpha_{ij} = 0\).
After that, the exponents \(\alpha_{ij}\) are recalculated using \eqref{eqn:softmax} and the simplification rules in Sec.~\ref{sec:DC} are applied.
The process of pruning extracts the factor node structure which was learned via the exponents \(\alpha_{ij}\).

\begin{figure}[t]
    \centering
    \includegraphics{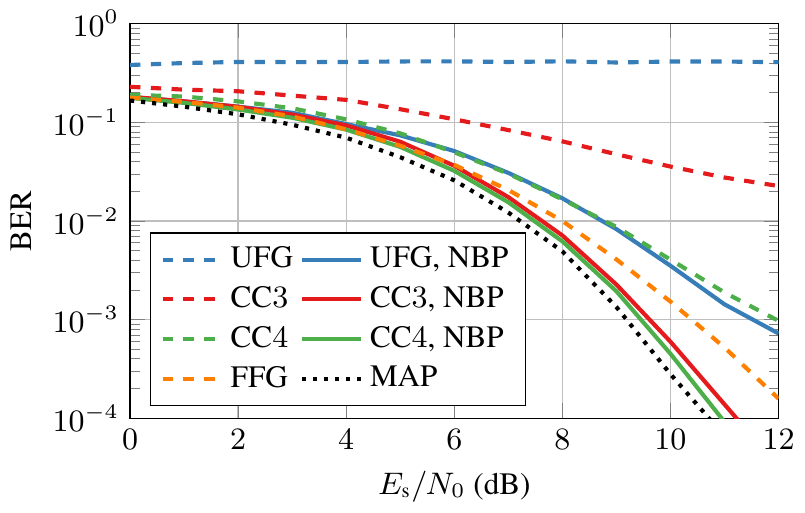}
    \caption{BER performance of the symbol detection using the SPA with several factor graph models. CC\(3\) and CC\(4\) have been learned using containers with degree \(3\) and \(4\), respectively. We used \(10\) SPA iterations for every factor graph model except for CC\(3\), where \(7\) iterations were used because more iterations resulted in a performance loss.}
    \label{plot:results}
    \vspace{-0.4cm}
\end{figure}

Since the structure is optimized by continuous parameters, CC can be easily combined with neural belief propagation (NBP)~\cite{nachmaniLearning2016}. NBP adds individual weights for each message and each iteration in the SPA. These weights are used to mitigate the negative effect of short cycles. This method was analyzed in \cite{schmidLowcomplexity2022} for the UFG and yields significant performance gains (see Fig.~\ref{plot:results}). In this work, we train these weights simultaneously with the structure of the factor graph, i.e., \(\vec{\beta}_i\), where the NBP weights are initialized with \(1\).

\vspace{-0.1cm}
\section{Results}\label{sec:results}
\vspace{-0.1cm}
First, we analyze the performance of CC without NBP and pruning for factor graphs with container degrees \(3\) and \(4\). We refer to the results as CC\(3\) and CC\(4\), respectively. 
The performance of the learned factor graphs is shown in Fig.~\ref{plot:results}.
The BER of the optimized models lies between the UFG and FFG.
Therefore, the CC models can be interpreted as an interpolation between both models as expected before.
The results for CC with NBP are also shown in Fig.~\ref{plot:results} (``CC\((3/4)\), NBP''). In combination with NBP, both the CC\(3\) and CC\(4\) model achieve near-MAP performance.

\begin{figure}[t]
    \centering
    \begin{minipage}{57mm}
        \includegraphics{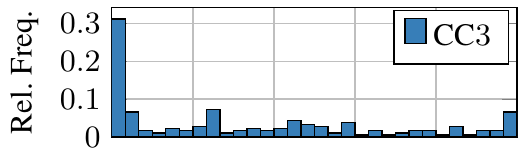}
        \includegraphics{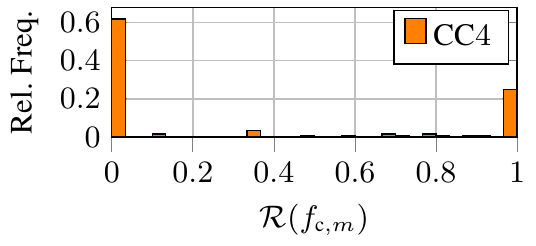}
    \end{minipage}%
    \hfill
    \begin{minipage}{27mm}
        \includegraphics{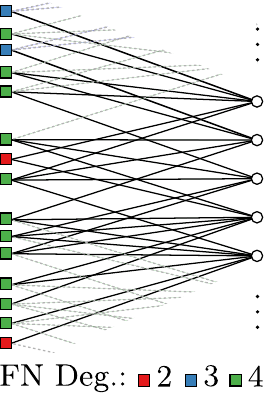}
    \end{minipage}
    
    \caption{Left: Histogram showing the relative frequencies of the relevances of all FN containers in the optimized factor graph CC\(3\) and CC\(4\). Right: Section of CC\(4\) after pruning showing the first \(5\) VNs and their connected FNs.
    }
    \label{plot:Pruning}
    \vspace{-0.2cm}
\end{figure}

To analyze the pruning performance, we calculated the
\emph{relevance}
\(
    \mathcal{R}(f_{\text{c}, m}) \coloneqq
    \max_{(i, j) \in \mathcal{M}_m} 
    \alpha_{ij}
\)
of each FN container \(f_{\text{c}, m}\) for both optimized models CC\(3\) and CC\(4\).
A low relevance indicates that the container has only a negligible impact on the SPA and can be pruned. The relative frequencies of the relevances are shown in Fig.~\ref{plot:Pruning}. The histogram shows that a relatively large proportion of containers (around \SI{30}{\percent} or \SI{60}{\percent}) is irrelevant.
This is possibly due to the fact that fewer FNs lead to fewer cycles that negatively affect the performance.
This effect allows the pruning of these negligible containers without performance loss:
A parameter analysis showed that a pruning threshold \(\alpha_{\idx{thr}} = \num{0.01}\) leads to no noticeable performance loss at \SI{10}{\dB} but reduces the complexity of the factor graph significantly as shown in Tab.~\ref{tab:distribution-fn-degree}.
\begin{table}
    \caption{Distribution of the degree of the pruned FN containers.}
    \centering
    \begin{tabular}{lrrrrr}
        \toprule
        Degree  & \(1\) & \(2\) & \(3\) & \(4\) & Pruned \\
        \midrule
        CC\(3\)  & \SI{0}{\percent} & \SI{30}{\percent} & \SI{50}{\percent} & - & \SI{21}{\percent} \\
        CC\(4\) & \SI{0}{\percent} & \SI{5}{\percent} & \SI{4}{\percent} & \SI{30}{\percent} & \SI{61}{\percent} \\
        \bottomrule
    \end{tabular}
    \label{tab:distribution-fn-degree}
    \vspace{-0.2cm}
\end{table}
Figure~\ref{plot:Pruning} visualizes the learned factor graph for CC\(4\) after pruning.
Iterative pruning and retraining of the factor graph structure as in~\cite{buchberger2020pruning} is subject of future work.

\vspace{-0.15cm}
\section{Conclusion}
We presented a machine learning framework to learn the structure of a factor graph in an end-to-end manner. The method is applied to symbol detection on inter-symbol interference channels as example inference task. The structure of the factor graph is learned by clustering FNs using optimized continuous weights. Our numerical evaluations show that the learned structures interpolate the performance and complexity between the UFG and the FFG. In combination with NBP, we achieve near-MAP performance for an exemplary channel model.
Subject of future work is the analysis of this method for other factor graph applications.

\end{document}